\documentclass[journal,letterpaper,compsoc]{IEEEtran}

\ifCLASSINFOpdf
  \usepackage[pdftex]{graphicx}
\else

  \usepackage[dvips]{graphicx}
\fi

\usepackage{url}
\usepackage[nocompress]{cite}
\usepackage{amsmath}
\usepackage{color,soul}
\usepackage{subfigure}
\usepackage{epsfig}
\usepackage{epstopdf}
\usepackage{enumerate}

\begin{document}

\title{Fast Cell Discovery in mm-wave 5G Networks with Context Information}

%Obstacle Avoidance Cell Discovery using mm-waves Directive Antennas in 5G Networks
%Context Information for Fast Cell Discovery in mm-wave 5G Networks

%Towards an Appropriate Beamforming Scheme for Initial Cell Discovery in mmW 5G Cellular Networks
%Comparative analysis of initial access techniques in 5G mmwave cellular networks
%Context Information Based Initial Cell Search for Millimeter Wave 5G Cellular Networks
%Directional cell discovery in millimeter wave cellular networks
%Directional cell search for millimeter wave cellular systems

\author{Ilario Filippini,~\IEEEmembership{Senior Member,~IEEE,}
        Vincenzo Sciancalepore,~\IEEEmembership{Member,~IEEE,} Francesco Devoti, and\\
		Antonio Capone,~\IEEEmembership{Senior Member,~IEEE}% <-this % stops a space
\IEEEcompsocitemizethanks{\IEEEcompsocthanksitem Ilario Filippini, Francesco Devoti and Antonio Capone are with Politecnico di Milano, Dipartimento di Elettronica, Informazione e Bioingegneria, Piazza L. da Vinci, 32, 20133 Milan, Italy. E-mail: \{name.surname\}@polimi.it. 
%\protect\\
\IEEEcompsocthanksitem Vincenzo Sciancalepore is with NEC Europe Ltd., Heidelberg, Germany.  E-mail: vincenzo.sciancalepore@neclab.eu.

}% <-this % stops an unwanted space
}

\markboth{Submitted to IEEE Transactions on Mobile Computing }%
{}

\IEEEtitleabstractindextext{%
\begin{abstract}
The exploitation of mm-wave bands is one of the key-enabler for 5G mobile radio networks. 
However, the introduction of mm-wave technologies in cellular networks is not straightforward due to harsh propagation conditions that limit the mm-wave access availability. Mm-wave technologies require high-gain antenna systems to compensate for high path loss and limited power. As a consequence, directional transmissions must be used for cell discovery and synchronization processes: this can lead to a non-negligible access delay caused by the exploration of the cell area with multiple transmissions along different directions.

The integration of mm-wave technologies and conventional wireless access networks with the objective of speeding up the cell search process requires new 5G network architectural solutions. Such architectures introduce a functional split between C-plane and U-plane, thereby guaranteeing the availability of a reliable signaling channel through conventional wireless technologies that provides the opportunity to collect useful context information from the network edge.

In this article, we leverage the context information related to user positions to improve the directional cell discovery process. We investigate fundamental trade-offs of this process and the effects of the context information accuracy on the overall system performance. We also cope with obstacle obstructions in the cell area and propose an approach based on a geo-located context database where information gathered over time is stored to guide future searches. Analytic models and numerical results are provided to validate proposed strategies. 
 
\end{abstract}

}

\maketitle
\IEEEpeerreviewmaketitle

\IEEEraisesectionheading{\section{Introduction}\label{sec:introduction}}

\thispagestyle{empty}	%remove number on the first page

% !TEX root = main.tex
In this phase of the process that will lead to the definition of the 5th generation (5G) of wireless access networks, the mobile industry sector is facing a traffic demand revolution that is strongly impacting on the wireless network performance. Guaranteeing high peak rates and low latency is no longer an option  only for few dedicated customers: advanced solutions must be developped to cope with smartphone and tablet applications' requests, which are pushing operator networks to face wireless resource scarcity. In this context, the compelling availability 
%of the currently used frequency bands 
of new big spectrum shares is drawing the attention of academic and industrial partners 
%with the aim of exploring 
to explore frequencies above the conventional 5GHz. To this extent, a promising opportunity is represented by millimeter wave (mm-wave) communications, %at 30-60 GHz. 
%In this emerging context, the advantage of exploiting mm-wave is 
which give a two-fold advantage: i) additional unlicensed spectrum bands can be smartly turned to good account, ii) mm-wave spectrum provides extremely wide bandwidths (up to 1GHz), enabling ultra-high data rates for greedy users \cite{survey_rappaport}.

Having been used for many years as a reference technology in point-to-point links, standardization groups are now looking at mm-waves in the novel perspective of the 5G network design, which enables their use in the radio access network (RAN).
%the perspective of enabling their use in the access part of the network, in the view of 5G network design. 
%This is certainly a big challenge, however mm-wave communications is one of the main players called to boost the network capacity and dramatically increase the data rate available for end users. 
This opens up new challenges as such a new solution for advanced RAN technologies suffers from 
%While this new Although this new application could seem straightforward, 
severe propagation losses combined with an adverse propagation environment. On the one hand, the weak ability to diffract around obstacles makes mm-wave communications vulnerable to blockage issues. Even the characterization of the propagation conditions is a challenging task, due to the large number of different aspects that can influence the signal \cite{HHI}, which must be carefully addressed.
%mm-wave propagation is characterized by very high attenuation that limits achievable ranges, indeed, path loss increases with the square of the frequency. 
%Moreover, most of the objects appear as opaque obstacles and cannot be penetrated by signals. The characterization of the propagation conditions is actually a challenge by itself, due to the number of different parameters that can influence the signal, and the very diverse physical scenarios to be considered~\cite{HHI}.
%
On the other hand, the short wavelength allows the use of antenna arrays with a relatively large number of elements that can be accommodated in a small space, both on the base station (BS) and on the user's mobile terminal (MT) \footnote{Within the text we use the terms \emph{user} and \emph{mobile terminal} interchangeably.}. Therefore, the research trend in mm-wave transmission technologies is currently moving towards the use of advanced beamforming techniques to concentrate the overall radiated power on very small angles, so as to extend the transmission range (i.e., the cell coverage) and to easily track terminals while they move within the coverage area. These techniques bring mm-wave small cells to cover up to a few hundreds of meters~\cite{mmwave_eLAN,coverage_analysis}.
%
%allow to design small BSs for mm-wave cells with a coverage area up to a few hundreds of meters~\cite{mmwave_eLAN,coverage_analysis}.

\begin{figure}[t!]
\centering
\vspace{-1cm}
\includegraphics[width=\columnwidth]{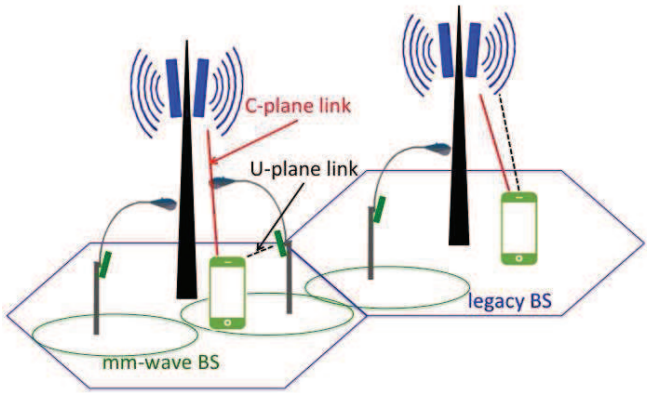}
\vspace{-1.2cm}
\caption{Example of C-/U-plane split architecture in mm-wave Radio Access Networks.}
\label{fig:arch}
\end{figure}

Nonetheless, the integration of mm-wave small cells in the current legacy RAN architectures exhibits its major limitations. 
%the use of mm-wave BSs in legacy cellular architectures is not straightforward, the peculiar characteristics of the mm-wave propagation requires a careful revision of current network access practices. 
Even resorting to the most advanced beamforming techniques, a full mm-wave small-cell deployment presents critical service disruptions and signal losses due to the high attenuation and the presence of obstacles, e.g.,  including human bodies, which can seriously hamper the signal propagation. 
To overcome this issue, we rely on a promising cellular architecture based on heterogeneous layers that blends together the emerging SDN paradigm and the functional split between the user plane (U-plane) and control plane (C-plane)\,\footnote{Although such a functional separation is not a novel concept in the network design, it has been recently introduced in the mobile cellular area. Beyond Cellular Green Generation (BCG$^2$) \cite{bcgwons} project within the GreenTouch Consortium (\url{http://www.greentouch.org}) has proposed the first complete architecture in 2011 for energy-saving purposes.}. While the former enables network controllers to centrally perform network operations, such as resource allocation, beam steering, etc.~\cite{SDN-mmwaves}, the latter guarantees a full coverage by exchanging signaling messages with legacy BSs (typically macro cells) and provides ultra-high capacity channels by means of mm-wave small cells~\cite{split} activated on-demand, as shown in Fig.~\ref{fig:arch}. 
%Mm-wave data connections can be made available in areas of particular interest, e.g., areas with important social events or with high traffic concentration.
Capacity layers are added to provide high transmission rates and low latency around hot spots, e.g., areas with important social events or with high traffic concentration. Mobile terminals are thus characterized by a double connectivity: one permanent connection is established to manage signaling, while a data bearer is activated on-demand for high-rate data transfers.
This is the solution currently considered within the MiWEBA\footnote{FP7-ICT EU-Japan Millimetre-Wave Evolution for Backhaul and Access (MiWEBA) project, \url{http://www.miweba.eu}.} project and other main projects on 5G (e.g., MiwaveS\footnote{FP7-ICT EU Beyond 2020 Heterogeneous Wireless Network with Millimeter Wave Small Cell Access and Backhauling (MiWaveS) project, \url{http://www.miwaves.eu}.}, 5grEEn\footnote{EIT-ICT Towards Green 5G Mobile Networks (5grEEn), \url{http://wireless.kth.se/5green}.}). Additionally, it also paved the road for a new network concept where network control functions are virtualized and optimally placed within the core network~\cite{NFV-specs}.

%This novel network design has to address several aspects when 
Although mm-wave transmissions promise to revolutionize the wireless access experience, considering mm-wave technologies as access network facilities brings additional complexity in network operations, e.g., the initial network access.
%Mm-wave technologies involve several have a disruptive effect on access network functions, they provide huge data rates, but they make network operations more complex. 
%Indeed, even the simple network access must be carefully investigated. 
Differently from previous specifications wherein synchronization signals were typically broadcast within the cell area and mobile terminals sensed and demodulated the synchronization signal, the directional nature of mm-wave communications requires the transmitter and the receiver to be spatially aligned
%Line-of-Sight (LOS) (or quasi-LOS) transmission 
, and thus, a discovery process must be executed.
%
%transmitters and receivers must beam each other before a proper connection can be established. 
This process consists in a geographical scan of the cell area (by both transmitter and receiver), sweeping through all possible directions.
% until they may beam each other. 
When the discovery process is not properly designed with smart mechanisms and algorithms, a large delay might be introduced in the user access, negatively impacting handover procedures~\cite{li2013anchor}, as well as the user quality of service (QoS). This becomes especially critical in scenarios where mm-wave communications are used in an opportunistic manner, by exploiting their extremely high data transfer rate in occasional short link activations.

%This procedure can delay the access to the network, which is critical when using small cells in an opportunistic manner. Clearly, this also negatively impact handover procedures \cite{li2013anchor}.
%

In this article, we present a thorough study on smart cell discovery schemes by highlighting advantages and drawbacks. Our analysis unveils a \emph{novel fast and efficient cell discovery procedure} relying on the new architecture proposal. In particular, driving the initial search on the basis of the \emph{user-location information} collected from the separate C-plane signaling, our solution outperforms the currently available discovery procedures by limiting the access delay.
We develop an \emph{accurate analytic model} taking into account directivity effects on both BS and MT side in order to predict the performance and to compare different discovery algorithms. Analytical results are further supported by exhaustive simulations to assess the impact of the uncertainty in the localization and propagation impairments in realistic scenarios. 
%
%We assume an external localization service providing an estimate of the user position to the network. Once such information is acquired, the discovery process is truly guided to synchronize the user. However, 
%We address the uncertainty in localization accuracy and the propagation impairments that might differ the entire process investigating how they impact of different cell discovery approaches.
%while the discovery procedure explores a number of beamforming configurations considering the probability of being revealed by the user. 
%a new architecture for enabling mm-wave communications in wireless access networks, discussing several advantages and highlighting technical challenges. 
%Among the latter, on the of most important is the cell discovery for the initial access. To this extent, 
%
%
%A performance evaluation of those approaches has been conducted by extensively simulating their behaviour in realistic scenarios. 
%in order to get useful insights on the several non-intuitive trade-offs. 
%In addition, numerical results have been verified with results from an analytic model designed to predict the discovery time performance according to the user distribution.
%
To the best of our knowledge, this is the first work investigating the \emph{impact of obstacles} on the cell search process. Their effect, combined with the inaccuracy of the location information, can severely harm the cell discovery procedure. To deal with this issue, we propose a \emph{geo-located context database} able to speed up the cellular attachment operations by storing and processing the information about the previous cell discovery attempts. The proposed framework enables the use of mm-wave communications in radio access networks by providing a fast and reliable cell discovery procedure in realistic network scenarios.
%that, storing the information on the previous cell discovery attempts, can speed up the cell acquisition process.

The article is organized as follows. We address the mm-wave initial \emph{cell discovery} problem in Section~\ref{sec:cell_disc}. In Section~\ref{sec:related}, we discuss related works on directional cell discovery. We describe the proposed cell discovery algorithms and their analytic models in Section~\ref{sec:obstaclefree}, while a thorough analysis of the numerical results is carried out in Section~\ref{sec:obstaclefree_results}. The effect of the obstacles and the impact of the geo-located context database are described in Section~\ref{sec:obstacleaware}. Finally, in Section~\ref{sec:implementability}, we discuss implementability issues of the proposed approaches. Section~\ref{sec:concl} ends the articles with some concluding remarks.

\section{mm-Wave Initial Cell Discovery}
\label{sec:cell_disc}
% !TEX root = main.tex

%The cell discovery process in mm-wave cells requires to directionally transmit signaling messages. It proceeds through scanning different beam directions in order to explore the entire service area. Upon beaming a joining user, tracking and tracing procedures allow the mm-wave small cell to start the association phase and, then, begin a data exchange. 

Technical challenges must be taken into account when applying a functional split to the cellular network. While legacy control functions assume that the service requests are collected and served by the same network entity, e.g., a base station, the functional separation abstracts the resource management functions by granting a full network view to an independent network entity.
This requires a richer user-context information in order to properly activate data network elements and, in turn, to approach the optimal utilization of system resources. As a consequence, the resource management becomes more complex, and thus, sophisticated resource allocation algorithms are required. 

Despite the functional split, some of the low level control functions, which are strictly related to the physical layer, cannot be delegated to a separated signaling connection with a different base station. Among these functions, the cell discovery process in mm-wave RAN is critical as it requires an initial directional scan that introduces an access delay.

In this article we propose fast cell discovery algorithms for mm-wave RAN that can smartly explore the small cell area while coping with access delay issues. They are based on the user-context information, retrieved through the separated signaling connection established by means of conventional cellular technologies. We show that even coarse information can greatly improve the cell discovery performance.

\begin{figure} [!t]
\centering
	\includegraphics[width=0.4\textwidth]{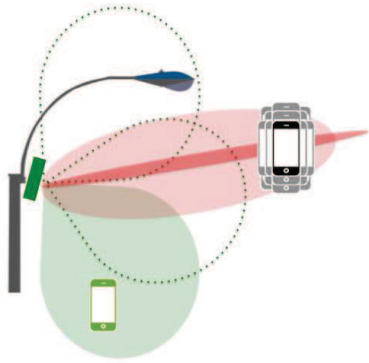}
	\caption{\footnotesize Directional cell discovery examples. }
	\label{fig:dir_search}
	\vspace{-1.5em}
\end{figure}

Let us consider a simple cell discovery strategy consisting in sweeping through all possible antenna configurations while looking for a rendezvous between the mm-wave BS and the user. This process might take long time so that the synchronization phase might be severely delayed. Current antenna technologies allow to use many different antenna configurations \cite{raghavan2016beamforming}: \emph{different beam widths} can be available, and, according to the beam width, many \emph{different pointing directions} can be used \footnote{In this work, we assume a fixed transmission power for the mm-waves small cell deployment following the current standardization bodies (e.g., 3GPP or NGMN). However, our proposal could be easily extended to admit advanced power control schemes.}.

The beam width selection has a significant impact on the discovery procedure. Large beam widths allow to scan faster over the surrounding space --fewer switches--, but they can reach relatively close users. Vice versa, narrow beams can cover far away users, but they require a large number of antenna configuration changes to scan the entire space. As a consequence, the discovery procedure must undergo a trade-off between speed and range of the space exploration. The best mix clearly depends on available antenna configurations and users distribution. As shown in Fig.~\ref{fig:dir_search}, the green wider beam allows to explore the space with a few configuration switches, however, it can cover only the green user. Indeed, the black user, farther away, cannot be reached with the green beam, but rather with a thinner beam, such as the red one. In this case, several switches are needed to cover the same angular span so as to eventually find the user.

The context information is another relevant factor that should be considered. Within a split architectural paradigm, the cell discovery might take advantage of acquiring information on mm-wave BSs and MTs. Generally speaking, the initial access procedure can be improved by richer information, e.g., terminal positions, channel gain predictions, user spatial distribution, antenna configurations successfully used in previous accesses, and so on. Ideally, if the context provided perfect information, mm-wave BS and user could directly discover and point each other with a narrow beam in one single step. In practice, inaccuracy is unavoidable, hence, context information can only be used to narrow down the search space and limit the searching time. Therefore, an advanced discovery algorithm is crucial to guide the search through the most likely successful configurations trying to minimize the number of attempts, even when only inaccurate context information can be obtained. 
%The quality of the context information may influence the cell discovery algorithm. 
For instance, Fig.~\ref{fig:dir_search} shows a black user affected by a unprecise position information. 
%\oldtext{The choices of the discovery algorithm could range from selecting a very narrow and long range beam to increase the gain toward the direction of the user estimated position, to choosing a large beam width in order to reach the user with lower power but being more robust to user location errors.}
The choices of the discovery algorithm could range from selecting a very narrow and long range beam --so as to increase the gain toward the user's estimated position-- to choosing a large beam width to be more robust against user location errors at the cost of a reduction in the user received power.

\begin{figure} [!t]
\centering
	\includegraphics[width=0.4\textwidth]{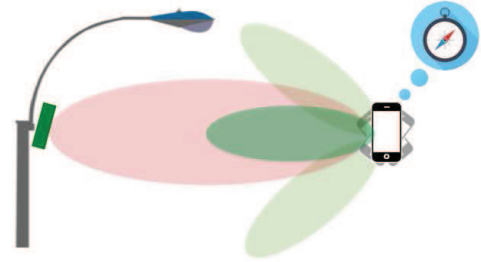}
	\caption{\footnotesize Directionality issues at mobile terminal side. }
	\label{fig:dir_ms}
	\vspace{-1.5em}
\end{figure}

User's location plays a relevant role in speeding up the discovery when the context information is considered. Given the knowledge about the network deployment, the BS exploits user's location to point its beams towards the network area where a user is expected to be found. However, the same assumption cannot be made at user's side. Indeed, both position and orientation can only be estimated. In particular, the orientation is extremely critical as a small error can seriously impair every context information benefit, as shown in Fig.~\ref{fig:dir_ms}. A rotation of the MT antenna array can lead the receiver to focus towards wrong directions, because of a different reference system. 
%\oldtext{Moreover, the limited availability of energy and computational resources at mobile terminal with respect to BS reduces the complexity of implementable discovery algorithms.} \addcommentblue{
Moreover, the limited energy and the limited computational power available at MT with respect to BS, significantly reduces the complexity of implementable discovery algorithms. Therefore, it is reasonable to expect basic strategies to address directionality at user side.

\begin{figure} [!t]
\centering
	\includegraphics[width=0.4\textwidth]{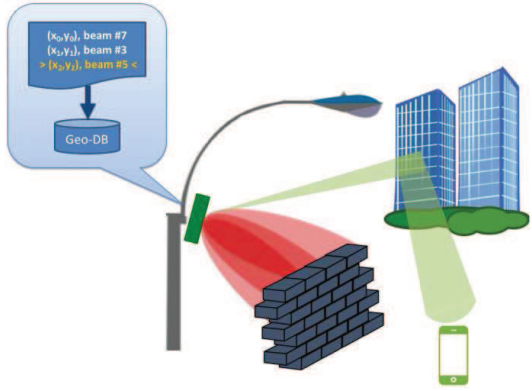}
	\caption{\footnotesize Using geo-located context database to manage reflected paths. }
	\label{fig:obstacle}
	\vspace{-5mm}
\end{figure}

Finally, due to unavoidable severe shadowing effects, obstacles (with static or quasi-static behavior) may strongly affect the cell discovery performance.
%
% issue to be addressed in mm-wave access is the strong fading effect caused by obstacles. 
On the one side, due to the high frequency, every object (i.e., human bodies, cars, trees, furniture, etc.) behaves as an opaque body, thus preventing signals from propagating through it. On the other side, the mm-wave propagation channel has a quasi-optical nature. Flat surfaces can be considered as mirrors~\cite{rappaport2012angle}, making reflected paths characterized by a non-negligible received signal strength, i.e., a viable alternative to the direct path, as supported by the experimental measurements conducted in~\cite{weiler2016millimeter}. %\oldtext{Hence, spatial scheduling solutions may even reach hidden users and provide ultra-capacity connectivity, if correctly designed to leverage reflected paths to overcome obstacles, as show in Fig.}
Hence, spatial scheduling solutions, if properly designed, may even reach hidden users and provide them ultra-capacity connectivity by leveraging reflected paths to overcome obstacle obstructions, as show in Fig.~\ref{fig:obstacle}.
%reflected paths, instead of Line-Of-Sight (LOS) paths, can be used to overcome obstacles along the direct path and easily reach users, as show in Fig.~\ref{fig:obstacle}. This is fully support Indeed, reflected paths are characterized by signal streghts comparable with the directed path, as confirmed by experimental measurements in~\cite{weiler2016millimeter}. 
%

Please note that the use of reflected paths may reduce the efficiency of basic cell discovery algorithms relying on position information, as they typically assume the LOS beam availability. When the LOS beam is obstructed, the algorithm must search among other antenna configurations until synchronization operations can be successfully performed. This makes the initial cell discovery process even more challenging. Nevertheless, storing the history of past attempts per user position in a context database can substantially help to derive information on the most likely configurations for a specific location. Approaching the search with the previous successful antenna configuration, and then, in case of failure, switching to a smart discovery algorithm, could be a simple but effective solution that greatly reduces the discovery time. The management of this geo-located context database is an aspect that must be carefully investigated, as it can provide a remarkable speed-up to the cell discovery process in mm-wave RAN facilities.

Although in our study we focus on the initial access, when the user joins the network, whenever a handover procedure occurs (or, more generally, the user connection is dropped) a new cell discovery procedure must be instantiated in order to provide mm-wave capabilities to the user (following the same procedure used for the initial access). Differently, an alternative approach could be adopted during transitions from and to an idle mode: delaying the mm-wave channel release could help to re-establish a new connection thanks to beam tracking functions. However, lower access delays are achieved at expenses of higher energy consumption levels. We leave this analysis as a future research direction and out of the scope of this article.

In the rest of our work, we start considering an obstacle-free scenario wherein we propose smart discovery algorithms. They provide advanced procedures to minimize the number of antenna configuration attempts in order to establish a connection with a signal power level sufficient to acquire the synchronization signal. After a complete dissertation, through an analytical and a numerical analysis of the obstacle-free case, we introduce obstacles in our scenarios. We discuss how context information about past successful attempts could be processed and used to further improve the overall system performance.

\section{Related works}
\label{sec:related}
% !TEX root = main.tex

Mm-wave technologies rely on high-gain directional antennas. While they are playing a leading role in the design of future networks, their issues and challenges have been extensively studied in the past for ad-hoc wireless networks~\cite{jakllari2007integrated}, where MAC protocols have been introduced to cope with the initial access problem. In~\cite{choudhury2006on}, protocol details are provided with a focus on the association and discovery phase in short-coverage environments, whereas Ramanathan \emph{et al.} present in \cite{Ramanathan2005ad} a complete framework for ad-hoc networks adopting directional antennas, pointing out technology challenges and off-the-shell solutions.
A widely-accepted conclusion of all these works is that, while improving spatial reuse and extending the link range, directional communications exacerbate deafness and hidden terminal problems and increase the complexity of the neighboring discovery process. Thus, most of the research works tend to consider a control plane management performed by omnidirectional interfaces (further discussed in Section~\ref{sec:implementability}) which orchestrates the beamforming operations~\cite{Korakis2008cdr-mac} or to assume a blind spatial search to detect the user signal~\cite{Choudhury2004deafness}. Interestingly enough, devices in Wireless Personal Area Networks (WPANs) are typically provided with omnidirectional sensing capabilities for neighboring discovery, whereas they can focus their beam towards incoming signals by using beam tracking algorithms, as shown in~\cite{wang2009beam}. Tracking algorithms to avoid obstacles on the direct path are used as well, as illustrated in~\cite{singh2009blockage}.

%WIGIG
The evolution of Wi-Fi networks has blended together recent mm-wave technologies and advanced directional MAC protocols. A new standard, IEEE 802.11ad, also known as WiGig, has been proposed to boost the WLAN performance by increasing the overall spectral efficiency. In this standard, the cell discovery process leverages the information collected during a periodic beamforming training period. Several solutions have been proposed to improve the standard.
A good solution is presented in~\cite{chandra2014adaptive} where Access Point (AP) beam parameters, such as its width, are optimally selected to reach every subset of surrounding users. Conversely, in the uplink context, the optimal directional beam information is analyzed in~\cite{chen2013directional}, where a novel MAC protocol selects the best direction found during a training period to beam towards a relay station whenever channel conditions degrade.
The work in \cite{beam_searching} presents a joint beam width selection and power allocation problem in order to maximize the overall network throughput in a mm-wave wireless network affected by deafness and misalignment impairments. Due to the huge complexity and the full topology awareness required by the optimal solution, the authors propose two standard compliant schemes which exploit directionality to maximize the reuse of available spectrum while imposing almost no computational load. Finally,~\cite{beam_steering_alg} proposes an interesting work on smart beam steering algorithms applied to mm-wave wireless LAN. Specifically, the proposal exploits previous information to narrow down the sector search space while users move.

None of the presented prior works aims at facing cell discovery issues in cellular networks, although they paved the way for developing new protocols to handle the ever-increasing requirements of future cellular networks.

%DISCOVERY PROCEDURE
Recent researches have started investigating mm-waves as a promising technology to be adopted in 5G networks. Some of the first works in this direction are~\cite{li2013anchor}, \cite{coverage_analysis}, \cite{capone2015context}, and \cite{capone2015_2}. The authors in \cite{li2013anchor} show that a mismatch between the discoverable network area and mm-wave service coverage could occur, and it is detrimental to the cellular network operations.
In~\cite{coverage_analysis}, an analytical model for cellular coverage (and rate performance) is presented by means of distance-dependent Line-Of-Sight probability function. Based on such a model, the authors have proven that dense mm-wave networks can achieve much higher spectral efficiency w.r.t. conventional cellular networks, while providing considerable cellular coverage. However, transmission blockages and coverage discontinuity make the full mm-wave deployment far from being practically viable. In \cite{capone2015context} and \cite{capone2015_2}, we initially proposed fast cell discovery procedures leveraging context information and provided a very preliminary approach to deal with obstacles and reflected rays.

In \cite{dir_cell_disc_rappaport}, the cell discovery procedure is addressed from a physical-layer perspective. Analysis and solutions are presented, however, the work focuses on acquiring channel synchronization, thus, it is somehow orthogonal to our proposal.
%Lastly, \cite{ew2015-CFS} presents an interesting cell search algorithm which optimally exploits the context-based information.
The authors in~\cite{mac_layer_perspective} discuss MAC layer design issues for cellular networks, identifying challenges and solutions. In addition, they provide a model for the spatial search delay when semi-directional or fully-directional mm-wave antennas are considered. Although being an accurate model, it assumes that directional synchronization pilots are transmitted in a first phase to get BSs and MTs synchronized both in time and space. Then, a spatial search is carried out to get BSs' and MTs' antennas aligned. This two-step access procedure might further delay the discovery process, therefore, we apply an one-step procedure in our proposal.
In~\cite{directional_access}, an evolution of the 3GPP standard for user association is presented by pointing out mm-wave challenges and, in particular, how to prevent connection losses. The authors propose two BS synchronization options: $i$) an omni-directional synchronization signal able to cover the entire cell area or $ii$) a sequence of narrow directional synchronization signals with fixed beam width, sequentially varying their directions. Conversely, in our work we analyze different sequences and combinations of beam widths and directions in order to shed light on the complexity of the cell discovery mechanisms and its delay implications.

An overview of iterative and exhaustive cell search schemes has been provided in \cite{zorzi1}, where also different algorithms have been compared. Very recently, in the work-in-progress paper \cite{zorzi2}, an approach based on context information and directivity of mobile antennas has been proposed. 

With respect to the above literature, in this article we provide the following novel contributions: 
\begin{itemize}
\item we propose smart cell discovery procedures relying on the C-/U-plane architectural split;
\item we propose a mathematical formulation to estimate the probability of being discovered within a short time window by using cell discovery procedures;
\item we introduce the use of mm-wave path reflections to avoid obstacles and reach hidden users;
\item we propose and evaluate a context-based approach which, leveraging past access information, minimizes future search delays;
\item we validate our approach by exhibiting outstanding results against basic search approaches in presence of obstacles, to the best of our knowledge, no advanced study has addressed this problem so far.
\end{itemize}

\section{Obstacle-free Fast Cell Discovery}
\label{sec:obstaclefree}
% !TEX root = main.tex

%Several solutions to directional neighbor discovery have been already proposed in other fields, such as ad hoc networks and personal area networks. However, none of those provides a scheme suitable for future 5G mm-wave network requirements, as relying on the availability of an omnidirectional pattern in some phases of the the discovery process.

Once a new user joins the network, it starts seeking for the synchronization signal sent through a BS beam. At one extreme, if no context information is available, the BS (user) randomly selects a beam width and a pointing direction while probing around to spot potential users (BSs). Upon receiving a signal power level above the minimum detection threshold, the user reports channel information and starts the association process. Clearly, this scheme might result in a very long cell discovery time due to the randomness of the user's position. 
%In addition, the user distribution heavily influences the discovery time, as the farther users, the thinner the beam-width, thus, the more beam directions must be explored.
At the other extreme, assuming a perfect knowledge of the user's position, the user's orientation, and the propagation environment, the BS readily computes the proper pair $s_0 = (w_0,d_0)$ (beam width, pointing direction) to provide a sufficient signal power level to the user, which gets connected in one single attempt. This trivially minimizes the cell discovery time.
The real condition is somehow between these two extremes: although not perfect, the information about the user position can significantly speed up the cell discovery procedure. 

Note that, an optimal beam selection implies the knowledge of the path loss between the MT and the BS, which could be estimated by using channel models \cite{HHI} or anchor-based prediction systems \cite{redondi2013context} with a given accuracy. However, for the sake of simplicity, given the statistical properties of such distributions, we can consider channel path loss and user location errors merged together in a unique equivalent location error. 

At BS side, if the user is not immediately detected with the initial beam $s_0$, due to the inaccuracy of the provided position, the discovery proceeds with other beams by means of a provided search sequence $s_1,...,s_n$.  
%We model the user-location uncertainty by considering the real user position distributed like a symmetric and independent bivariate normal distribution, as typically done in GPS systems.
As for the user side, we consider a simple discovery based on the SLS phase of IEEE 802.11ad BF protocol. We consider $M$ available beams that equally divide the full circle, they are sequentially explored until the BS beacon is correctly received. We assume a loose synchronization between MT and BS, which can be established through the legacy C-plane connection. At each BS beam switch, a full MT circular sweep is carried out, switching through its $M$ beams. When the MT has completed its switches, the BS can move to another beam and the MT sweep restarts. Note that in this process a BS beacon is transmitted at each MT switch.

When a MT and a BS activate two beams that allow them to point each other in a way that the BS beacon can be decoded at MT side, the cell discovery is concluded. This can be confirmed by the user via a prompt message transmitted through the mm-wave interface, or through the legacy C-plane interface that successfully notifies the BS. 
%the pair of BS and user beam identifiers can be transmitted via the  to acknowledge the successful connection. 
The best option relies on the network architecture and the mm-wave MAC protocol, which, although being an interesting issue, cannot be addressed in this article due to space limitation reasons. When the whole BS search sequence has been explored without a positive user feedback, the user is marked as unreachable.

%In order to boil down the analysis of the cell discovery problem to its essential aspects, we define a simplified setup. {\color{red} We focus on downlink transmissions of the synchronization signal, assuming an omnidirectional antenna at the user side, whereas a directional antenna is available at mm-wave BS \footnote{Note that this setup is currently considered in the SLS phase of IEEE 802.11ad BF protocol.}}. 

Let us consider a 2D environment (fixed elevation angle) and a steerable antenna, with a discrete set of beam widths $W_{-3dB}$ and pointing directions $D$.
The number of pointing directions with a beam width $w_{-3dB} \in W_{-3dB}$ is $N = 2\pi / w_{-3dB} $.
%\addcomment{Se non ho capito male $D$ e' il numero totale di pointing directions, quindi $N$ e' il numero di pointing directions dato un beamwidth, giusto? Se e' cosi' lo indicherei diversamente:\\}
%\addcommentblue{Given a beam width $w_{-3dB} \in W_{-3dB}$, the number of pointing directions is $N = 2\pi / w_{-3dB} $.}\\
%\addcomment{Mi ha confuso questa notazione: $\vert W_{-3dB} \vert$, che mi ha fatto pensare alla cardinalita' dell' insieme $W_{-3dB}$. }.
Beam width and antenna gain are linked by a Gaussian main lobe profile \cite{miwebaD51}:
\begin{equation}
\label{antenna_model}
\renewcommand{\arraystretch}{0.99}
\begin{array}{l}
G_{dB}(\phi,\theta) = \\ 10log\left(\frac{16\pi}{6.76\cdot w^{\phi}_{-3dB}\cdot w^{\theta}_{-3dB}}\right)-12\cdot\left(\frac{\phi}{w^{\phi}_{-3dB}}\right)-12\cdot\left(\frac{\theta}{w^{\theta}_{-3dB}}\right) 
\end{array}
\end{equation}
where $\phi$ and $\theta$ are defined as offsets between the main lobe direction and the elevation angle and azimuth angle, correspondingly, while $w^{\phi}_{-3dB}$ and $w^{\theta}_{-3dB}$ are respectively elevation and azimuth half-power beam widths. Since we consider only azimuth angles in the following, $w^{\theta}_{-3dB} = w_{-3dB}$.
%Fig.~\ref{fig:antenna_patterns} shows different antenna coverages for different beam-width values and highlights the width-vs-range trade-off.

%\begin{figure} [!t]
%\centering
%	\includegraphics[width=0.4\textwidth]{antenna_patterns_mod.png}
%	\caption{\footnotesize Antenna coverage for different beam-width values depicted on x-y axis, base stations are placed in the middle of the area (point $500$m, $500$m). Received power is averaged over a $10$x$10$m square area. }
%	\label{fig:antenna_patterns}
%	\vspace{-1.5em}
%\end{figure}

We cast the cell discovery problem into a search time minimization problem wherein, given a user behavior characterized by a sequential beam scan, a mm-wave BS must find $i)$ the beam width $w_{-3dB} \in W_{-3dB}$, and $ii)$ the pointing direction $d \in D$ in order to provide the user with a sufficient signal power level to detect the presence of the mm-wave BS, and thus, to synchronize. A cell discovery procedure must define the search sequence of beams $\mathbf{s} = s_0,...,s_n$ (i.e., \{beam width - direction\} pairs) to be sequentially activated by the mm-wave BS. %The sequence is explored until the synchronization signal can be received and successfully decoded by the user.

Clearly, the way $\mathbf{s}$ is designed strongly determines the performance of the discovery procedure. In the next section, we present an analytic model to predict the cell discovery time starting from a generic search sequence $\mathbf{s}$. Then, in Section~\ref{sec:obstaclefree_procedures} we discuss several design solutions for $\mathbf{s}$.

\subsection{Stochastic model}
\label{sec:obstaclefree_model}
% !TEX root = main.tex

%In order to model the mm-wave initial access, 
We consider a scenario where users are distributed, according to their nominal position (e.g., the one indicated by the localization system), following a bidimensional density function $f_{XY}$ in a square area of side length $l$. Each nominal position $(x,y)$ is affected by a localization error described by a symmetric and independent bivariate normal distribution characterized by parameter $\sigma$ and centered in $(x,y)$\footnote{We rely on the modelization of the localization error commonly applied into standard GPS systems~\cite{GPS_book}.}. User orientation is randomly chosen. Users are dropped in the obstacle-free square area and the mm-wave BS is placed in the middle.
\begin{figure}[t]
\centering
%\vspace{-0.5cm}
\includegraphics[width=0.7\columnwidth]{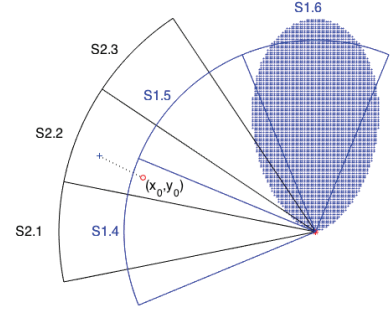}
\vspace{-0.5cm}
\caption{Beam footprint and model sector approximation. The dashed region is the coverage footprint approximated via sector S1.6. The red dot is user's nominal position, while the blue cross is its real position.}
\label{fig:beam_approx}
\end{figure}

As shown in the sector $S1.6$ of Fig.~\ref{fig:beam_approx}, we approximate the coverage footprint of a pair of beams selected by M and BS (i.e., the area surrounding the BS where a MT can correctly decode BS signals when the BS beam is active and the MT beam is pointing towards the BS) with a circular sector described by three parameters: radius, angular width, and pointing direction. We use the half-power beam width $w_{3dB}$ of the BS beam as sector \emph{angular width} and the same \emph{pointing direction} as of the BS beam, $p=k\cdot w_{3dB}$, with $k=0,..,\frac{2\pi}{w_{3dB}}-1$. As \emph{radius}, we consider the maximum distance, scaled by a factor $\eta$, at which user and BS can connect if both MT and BS beams are perfectly aligned. Basically, we consider the maximum antenna gain at both BS and MT.
Our design introduces two approximations: i) while the real footprints of adjacent beams slightly overlap, sectors with a fixed width can cover $2\pi$ radians without overlapping, ii) while the model considers a perfect MT-BS alignment, $M$ user beam directions are fixed, depending on user's orientation. Therefore, MT and BS beams could never be perfectly aligned during the MT scan. This reduces the maximum coverage distance, which we model with the factor $\eta$. However, these approximations allow to greatly simplify the model without substantially impairing its accuracy, as shown in the next section where model results are validated against simulated ones.

Given the user's nominal position $(x,y)$, the cell discovery starts with a BS beam that can potentially cover $(x,y)$, if the MT beam were pointed towards the BS. In the model, this translates into finding the circular sector with the largest width including $(x,y)$, which can be straightforwardly computed. Once the BS beam is activated, in turn, the MT sequentially activates its $M$ beams, starting from a random direction, until the BS beacon is detected.
%\addcomment{Scrivere prima che: i) il beacon della BS viene ripetuto ad ogni user beam switch, ii) beam discovery algorithm sono lato 
Please note that the BS beacon is retransmitted $M$ times on the same selected BS beam direction in order to help the synchronisation process from user side. %This assumption, together with the BS-MT loose syncronization one, allows us to state succesful beacon detection when the user's received signal power level is large enough to decode the beacon.
Given the low-computational resources on the user side, we only implement advanced beam discovery algorithms on the BS side whereas MT simply performs a sequential beam search. However, more-complex user searching algorithms may be addressed in future works. 

Wrong parameters in the initial BS search beam may result in an unsuccessful user connection, thereby forcing the cell discovery process to scan a sequence of BS beams, $\mathbf{s}$, determined by the search algorithm. If we label every possible BS beam configuration $(w_{3dB},p)$ with an identification number, we obtain that each search algorithm, $\alpha$, provided with the initial location $(x,y)$, defines a numerical sequence $\mathbf{s} = \mathbf{s}\left\{\alpha,(x,y)\right\}$ of beam identifiers. The $n$-th element of $\mathbf{s}$ indicates the circular sector corresponding to the BS beam selected at the $n$-th step of the search process at BS side. Once the MT is located within the coverage footprint, we assume that the connection will be established using the same BS beam and one of the $M$ MT beams.

In order to assess the performance of a cell discovery algorithm $\alpha$, we are interested in computing the probability of connection establishment at step $n$, at both BS and MT side. We name $p_{BS,n}^{\alpha}$ and $p_{MT,n}^{\alpha}$, respectively, the probability mass function (PMF) at BS and MT side. The PMF $p_{BS,n}^{\alpha}$ can be computed considering that a MT in the coverage footprint defined by the BS beam $s_n$ will be connected before the BS moves to the next beam. Therefore, it can be expressed as follows:
\begin{equation}
p_{BS,n}^{\alpha} = \iint p_{BS,n}^{\alpha}(x,y)f_{XY}(x,y)\text{d}x\text{d}y,
\end{equation}
where $p_{BS,n}^{\alpha}(x,y)$ is the conditioned probability of connection at the $n$-th step of algorithm $\alpha$, given the user's nominal location $(x,y)$. We can express the conditioned probability as:
\begin{equation}
p_{BS,n}^{\alpha}(x,y) = \iint_{A_n\left\{\alpha,(x,y)\right\}} \mathcal{N}(x,y;x,y,\sigma)\text{d}x\text{d}y
\label{eq:cond_find_prob}
\end{equation}
where the normal distribution of the location error $\mathcal{N}$ is described by the parameter $\sigma$, and shifted to $(x,y)$. Depending on the algorithm $\alpha$, $A_n\left\{\alpha,(x,y)\right\}$ is the area obtained as the difference between the area of sector $s_n$ and the area of the overlapping sector(s) with a smaller radius (i.e., larger width than $s_n$), which has (have) been already explored in the sequence $\left\{s_0..s_{n-1}\right\}$. If no overlapping sector belongs to the sequence, $A_n\left\{\alpha,(x,y)\right\}$ corresponds to the whole area of sector $s_n$. 

In Fig.~\ref{fig:beam_approx}, sector $S1.5$ overlaps only with sector $S2.3$, while sector $S2.2$ has two adjacent overlapping sectors, $S1.4$ and $S1.5$.
When $s_n = S2.2$ and only $S1.5$ is in $\left\{s_0..s_{n-1}\right\}$,
%\oldtext{$A_n\left\{\alpha,(x,y)\right\}$ splits into the annulus sector determined by $S2.2$ and $S1.5$, and the remaining circular sector obtained by subtracting the subsector containing $S1.5$ from $S2.2$}
%\addcommentblue{$A_n\left\{\alpha,(x,y)\right\}$ is determined by substracting the $S2.2$ and $S1.5$ areas intersection from the $S2.2$ area}.
$A_n\left\{\alpha,(x,y)\right\}$ splits into the annulus sector determined by $S2.2$ and $S1.5$, and the remaining circular sector obtained by substracting the overlapping area of $S2.2$ and $S1.5$ from the $S2.2$ area.
Since the definition of $A_n\left\{\alpha,(x,y)\right\}$ states that overlapping sectors are always wider than the overlapped ones, the case in which more than two adjacent sectors overlap with a narrower sector does not exist.
%\color{black}
%\addcomment{Non capisco: $S1.5$ e $S1.4$ sono entrambi sovrapposti a $S2.2$}

Finally, the definition of $A_n\left\{\alpha,(x,y)\right\}$ follows from the probability of connection at step $n$. Indeed, in order to have a successful discovery at step $n$, the user must lie in a region of the sector $s_n$ covered by none of the sectors $s_t$, $t \in \left[0,n-1\right]$, otherwise, the BS beacon would have been already detected before step $n$. The computation of $A_n\left\{\alpha,(x,y)\right\}$ is greatly simplified by the removal of the overlap among sectors with the same width.

As for the PMF at MT side, $p_{MT,n}^{\alpha}$, we can note that once the discovery probability at the corresponding BS step $\lfloor\frac{n}{M}\rfloor$ is computed, the probability of connection in one of the steps of the corresponding MT round is uniform and equal to $\frac{1}{M}$. Therefore, we can write:
\begin{equation}
p_{MT,n}^{\alpha} = \frac{1}{M} p_{BS,\lfloor\frac{n}{M}\rfloor}^{\alpha} 
\end{equation}

\subsection{Cell Discovery procedures}
\label{sec:obstaclefree_procedures}
% !TEX root = main.tex

%Interestingly, in this section we show how to cope with the cell discovery problem, focusing our attention on the discovery time, which may become very critical in the 5G network design. 

%We thoroughly approach the problem, carrying out a deep analysis on the key parameters presented in the previous section (i.e., beam-width $\theta_{-3dB}$ and pointing direction $d$). This sheds light on how to properly choose those values, in a very short time, in order to provide a sufficient signal level to incoming users to detect the BS presence. 

The sequence $\mathbf{s}$ of explored beams is the key-feature to determine the performance of the mm-wave cell discovery phase. Several approaches can be designed with different complexity, depending on the available user-context information. In this section, we focus on algorithms based on the mobile terminal position estimation, which is conveyed to the network through the separated C-plane within the initial service request.

\paragraph*{\underline{SLS}}A first procedure to populate $\mathbf{s}$ consists in a straightforward extension of the \textit{Sector Level Sweep} (SLS) phase of IEEE 802.11ad BF protocol. It considers a fixed beam width, set to the narrowest available value in order to preserve the largest coverage. The procedure starts by configuring the beam parameters pair $(w_0,d_0)$ (beam width and pointing direction) that cover the nominal user position $(x,y)$, then, it proceeds to circularly sweep through adjacent beams until the whole circle has been explored. Users that cannot be discovered by this procedure are defined as unreachable.

\begin{figure} [!t]
\centering
	\includegraphics[width=0.4\textwidth]{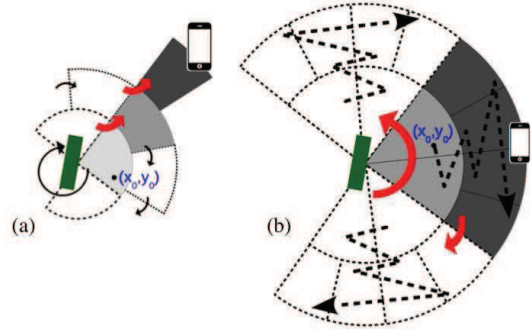}
	\vspace{-2.5em}
	\caption{\footnotesize Illustration of (a) Dynamic SLS and (b) Enhanced Discovery Procedure. Estimated user location is $(x_0,y_0)$.}
	\label{fig:algo}
	\vspace{-1.5em}
\end{figure}

\paragraph*{\underline{D-SLS}} As a first improvement over Sector Level Sweep, we apply a search paradigm that dynamically adapts beam widths, called \textit{Dynamic SLS} (D-SLS, see Fig.\ref{fig:algo}(a)). The search starts considering the widest beam, $(w_0,d_0)$, that can reach the user at $(x,y)$. If the user is not detected, the mm-wave BS sequentially scans around through every direction, keeping the same beam width $w_0$. If no user is still found, the mm-wave BS restarts the circular sweep with a reduced beam width. The next circular scan will cover a larger area and will consist of a larger number of attemps. The procedure repeats until every combination of beam width and pointing direction is explored. The rationale behind D-SLS is the idea of exploring the close-by neighborhood through wide beams thereby allowing to reach nearby users faster than SLS.
%to quickly explore the area close to the BS through wide beams, this allows to reach nearby users faster than in SLS. Far-away users require narrow beams and will be reached after the exploration of the nearby area.

\paragraph*{\underline{EDP}}In order to exploit the context information in a smarter way, we propose an enhanced version of the D-SLS procedure, named \textit{Enhanced Discovery Procedure} (EDP) (Fig.\ref{fig:algo}(b)). As in the previous approach, when a new user joins the network, the serving BS quickly computes the correct beam width and pointing direction, $(w_0,d_0)$, to properly beam the user based on the estimated position. If the position is not accurate, the user might not be spotted, thus, the BS scans the surrounding environment relying on $n$ circular sectors, each $(2\pi/n)$-radian wide. The first selected sector, $r$, corresponds to the area containing the estimated (inaccurate) position of the user. To overcome the inaccuracy of the position, the BS starts exploring the sector $r$ through beams with a fixed width $w_0$ and directions adjacent to $d_0$. During this exploration the BS considers alternate clockwise and counter-clockwise beam directions. If no user is reached, the BS narrows the beam width, points the beam again towards the estimated user position, and similarly explores adjacent beam directions within the sector. After completing the scan of the first sector without any user connection establishment, the same scan iteration is repeated for each of the other adjacent $n-1$ sectors, alternating clockwise and counter-clockwise sectors. In each sector, the exploration starts from the relative \{beam width, pointing direction\} pair corresponding to ($w_0$,$d_0$) in $r$. The process ends when the UE has been detected or when all $n$ sectors have been scanned without any user connection.

The Enhanced Discovery Procedure aims at providing a good trade-off between two opposite strategies: i) first scanning large azimuthal angles and then extending the range by narrowing the beam, or ii) first exploring narrow azimuthal angles until the maximum range is reached and then changing pointing direction. As shown in the next section, the most convenient strategy depends on UE distribution statistics and position accuracy.

\section{Numerical evaluation}
\label{sec:obstaclefree_results}
% !TEX root = main.tex

In this section, we assess the cell discovery performance of the proposed algorithms by means of numerical simulations carried out by an ad-hoc MATLAB\textregistered\, simulator. If not stated differently, we consider the following scenario. We place one mm-wave BS in the middle of a $450$ x $350m$ area, while 1000 users are dropped according to different distributions, as described later. We model the user-location uncertainty by considering the real user position distributed as a symmetric and independent bivariate normal distribution centered in the nominal position with parameter $\sigma_x = \sigma_y = \sigma$. We consider a location error $\epsilon = 3 \sigma$. The antenna gain is modeled with a Gaussian main lobe profile described in Eq.~\eqref{antenna_model}, while the path loss model used for the transmission is defined as follows:
\begin{equation}
\text{P}(l) = 82.02 + k\cdot 10 \log\left(\frac{l}{l_0}\right)
\end{equation}
where reference distance $l_0$ is $5$ meters and the propagation factor $k$ is $2.36$, if the distance between the transmitter and the receiver is larger than the reference distance, or $2.00$ otherwise. For more details, e.g., fading and other channel properties, we refer the reader to~\cite{miwebaD51}. The minimum signal level for primary synchronization signal (PSS) acquisition, $Th = -73dBm$, is directly derived from the empirical measurements presented in~\cite{dir_cell_disc_rappaport}, where a Signal-to-Noise-Ratio (SNR) greater than $10$dB has been experimentally confirmed. The BS transmitting power is set to $P_t=30dBm$.
% and kept fixed over a number of scenarios as downlink power control mechanisms introduce very small performance enhancements in mm-wave deployments at expenses of huge complexity. 
The smallest configurable beam width at BS side is $3$ degrees, which means the availability of non-overlapping beams pointing towards $120$ directions. Larger beam widths are obtained by proportionally reducing the number of directions to $72, 36, 24, 12, 8, 4, 3$, and $1$ for a total number of $280$ antenna configurations available at BS side. As for MT side, we consider a more limited hardware that consists in $M=12$ non-overlapping beams, with a beam width of $30$ degrees that can be pointed in $12$ fixed directions. Finally, if not stated differently, we measure the cell discovery time in terms of antenna configuration switches at MT side. This is directly related to the cell discovery time according to the antenna hardware specifications. In addition, note that we focus on the smart algorithms implemented only at BS side and their effects on the discovery performance. Indeed, we can consider the simple SLS-like algorithm at MT side as a given behavior, motivated by the limited amount of resources.

\subsubsection*{User distribution impact}
We first study the impact of the user distribution on the cell discovery time. We assume no context information about user geographical positions. The BS randomly chooses a beam width and a direction (i.e., the pair $\{w_{-3dB}$,$d\}$) from the set of all possible combinations: We called this scheme \textit{Random Discovery Algorithm}. Upon a choice, if no user is detected, the BS selects another pair of parameters ensuring that each pair is used once. At each BS switch, the MT performs a full scan of the surrounding area in $M$ sequential beam switches. If no user connection is established after probing all possible pairs of BS and MT antenna configurations, the user is marked as unreachable.

\begin{figure} [!t]
\centering
	\includegraphics[width=0.45\textwidth]{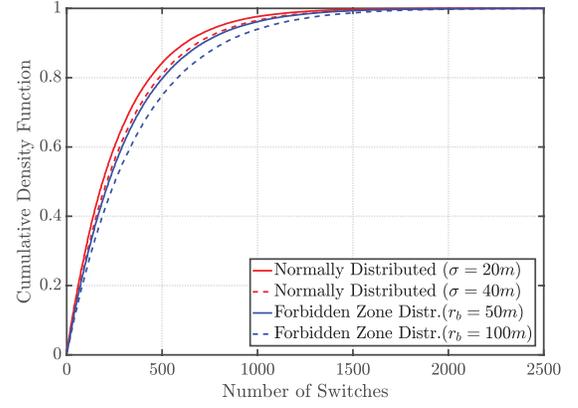}
	\caption{\footnotesize {CDF of the number of beam switches performed during the cell discovery phase considering different user distributions.}\color{black}}
	\label{fig:cdf_random_search}
	\vspace{-1.5em}
\end{figure}

\begin{figure}[h!]
\centering
	\subfigure[Population of $1000$ users normally distributed with $\sigma=10$.\color{black}]
	{
	\includegraphics[width=0.45\textwidth]{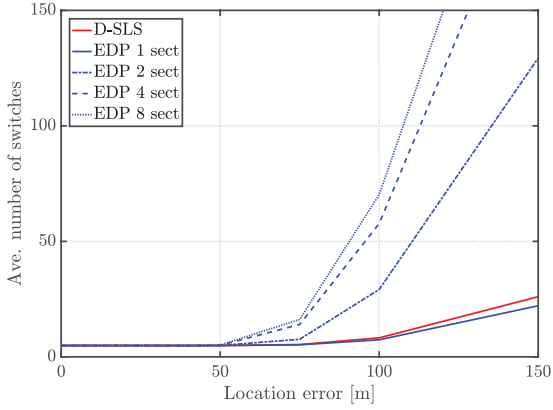}
	\label{fig:varySigma_10}
	}
	\subfigure[Population of $1000$ users normally distributed with $\sigma=100$.\color{black}]
	{
	\includegraphics[width=0.45\textwidth]{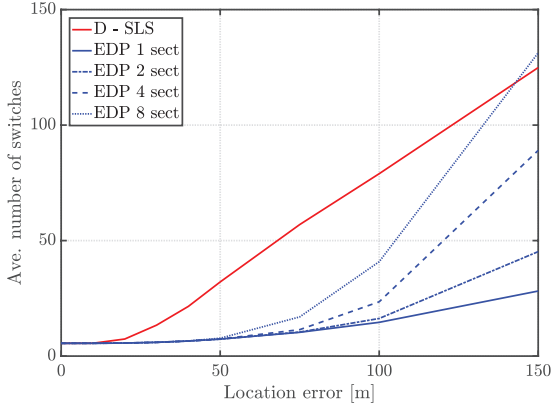}
	\label{fig:varySigma_100}
	}

	\caption{\footnotesize{Performance evaluation in terms of average number of beam switches required to establish a connection for different numbers of EDP sectors and localization error values.}\color{black} 
	}
	\label{fig:discovery_algorithms_comp}
\end{figure}

Fig.~\ref{fig:cdf_random_search} shows the cumulative distribution functions (CDF) of the number of beam switches required to detect incoming users. Different CDFs generated by $4$ different user distribution functions are compared in the plot. In the first curves, users are dropped in the area according to a 2D normal distribution centered in BS coordinates and with a standard deviation $\sigma$ equal to $20$ meters and $40$ meters, respectively. We consider the same distribution for the other two plots but with a small change: we include a forbidden zone of, respectively, $50$ and $100$ meters around the center of the cell, where users cannot be placed. This results in dropping all users on the cell-edge, which allows us to investigate the cell-border effect.

Results provide useful insights: User distributions that are denser nearby the BS experience a shorter delay. Vice versa, when the distribution is spread far from the BS, the cell discovery performance degrades, showing a high probability of presenting a large number of beam switches. Especially for these cell-edge scenarios, an improved cell discovery mechanism is needed.

\subsubsection*{Context: Geographical Position Accuracy}
In this section, we analyze the performance of the context-based algorithms proposed in Section~\ref{sec:obstaclefree_procedures} varying the accuracy levels of the position information. Fig.~\ref{fig:discovery_algorithms_comp} shows the average rendezvous time, in terms of number of antenna configuration switches, when the location accuracy varies and users are normally distributed over the rectangular area with different standard deviations $\sigma$. The figure shows curves for two algorithms: D-SLS, and EDP. For the EDP algorithm, four curves are plotted varying the number of sectors $n$ from $1$ to $8$. %\addcomment{Looking at SLS curve, it is clear that, while it could be a reasonable choice for small location errors, its performance quickly degrades when the error increases};
%\oldtext{D-SLS and EDP algorithms perform much better in this condition}
Due to the user location information, D-SLS and EDP algorithms outperforms SLS, which takes in average more than $500$ MT switches to establish a connection. Interestingly, considering EDP, going from one sector, $360$-degree wide, to more and narrower sectors makes the rendezvous time increase. Indeed, when the error increases, the higher error sensibility of narrower sectors negatively impacts on EDP performance. The algorithm is caught up in a deep exploration of a sector that could be wrong due to the position inaccuracy. The choice of a single $360$-degree-wide sectors seems to be the best choice for this scenario without obstacles.

When the user distribution concentrates users close to the BS, like in Fig.~\ref{fig:varySigma_10} with $\sigma=10$, the average number of required switches, using a single $360$-degree sector, decreases. This is due to the fact that users placed close to the BS can be covered with short-range (wide) beams and, consequently, the surrounding area can be scanned in few switches. However, Fig.~\ref{fig:varySigma_100}, where $\sigma=100$, reveals an interesting behavior when the EDP algorithm considers $45$-, $90$- and $180$-degree-wide sectors ($8$, $4$, and $2$ sectors): the rendezvous time decreases with respect to the results for concentrated users in Fig.~\ref{fig:varySigma_10}. The motivation is purely geometric: if a close user and a far-away user are characterized by the same location error, the error of the latter is seen through a narrow view angle, while the error affecting the former has a much larger impact on the view angle. This implies that location errors for users close to the BS, which are more when the normal distribution has a smaller $\sigma$, are magnified by angular-selective discovery procedures. 
Finally, EDP with $n=1$ performs better than D-SLS, as it prefers to activate beam configurations directed towards the user rather than to perform circular beam searches.

\begin{figure}[h!]
\centering
	\subfigure[Average number of BS switches.]
	{
	\includegraphics[width=0.45\textwidth]{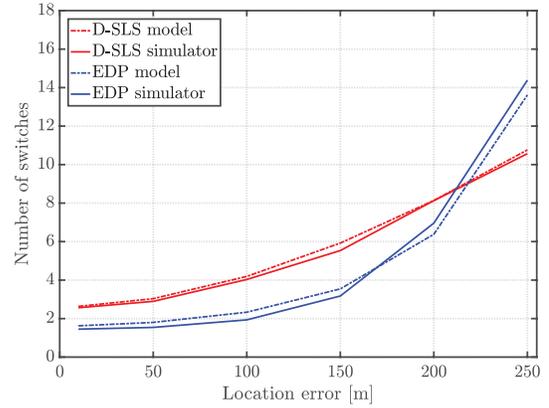}
	\label{fig:ave_nb_BSswitches}
	}
	\subfigure[Average number of MT switches.]
	{
	\includegraphics[width=0.45\textwidth]{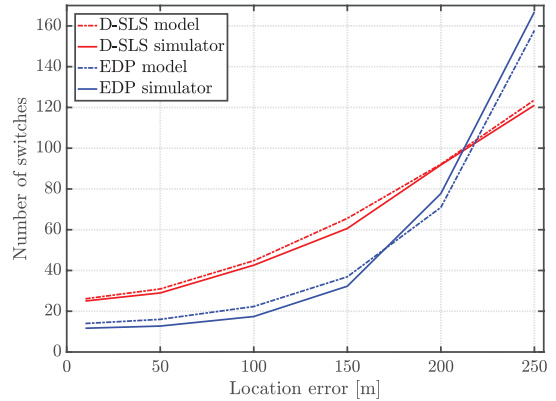}
	\label{fig:ave_nb_MTswitches}
	}

	\caption{\footnotesize{Comparison between analytical and simulated results at different location error values.}
	}
	\label{fig:model_accuracy}
\end{figure}

\subsubsection*{Model quality assessment}

In order to assess the level of approximation introduced by the model described in Section~\ref{sec:obstaclefree_model}, we have compared the performance predicted by the analytic model with those empirically obtained with the simulator. We have considered the same realistic scenarios described above, however, for sake of brevity, we show results only for the case of users dropped according to a uniform random distribution. Nevertheless, we want to remark that our proposed model works with any other user distributions, so as similar results can be obtained and similar conclusions can be drawn.

Fig.~\ref{fig:ave_nb_BSswitches} plots the average number of BS steps needed by EDP and D-SLS algorithms to establish a connection in case of different location errors. The figure confirms the high accuracy of the model by showing a very small gap between analytical and simulated values for every practical location error value. As expected, the figure shows that the number of required steps increases as the error increases. A similar behavior is shown in Fig.~\ref{fig:ave_nb_MTswitches}, where the number of switches at MT side is reported.

%Fig.~\ref{fig:pmf_comparison} compares the probability mass function (PMF) of the number of required steps for two different location error values, applying both D-SLS and EDP algorithms. When the error is small, most of the users are detected in $1$ step, as the beam to reach the nominal user position can cover, with very high probability, also the real position. As soon as the location error increases, PMFs shift towards larger values. Note that the D-SLS curve has a much longer tail that extends beyond the limits of the plot. In addition, D-SLS PMF has an oscillatory behavior. Indeed, due to monotone circular exploration, users are found with high probability at the beginning and at the end of the circle.

\section{Dealing with Obstacles}
\label{sec:obstacleaware}
% !TEX root = main.tex

Real environments are characterized by objects that can lie between mm-wave transmitters and receivers causing severe signal attenuations. These objects can be static (trees, buildings, etc.) or nomadic obstacles (human bodies, vehicles, etc.). While the effect of mobile obstacles can be modeled as a random fading and, thus, addressed by physical layer solutions, fixed obstacles systematically drop the signal and must be avoided from an upper layer perspective.

Since the propagation of mm-waves through obstacles is practically unfeasible, the only viable solution is to deviate from a LOS path and turn around the obstacle. Resorting to deviated paths is particularly efficient with mm-waves because of the good reflectivity of walls at high frequencies \cite{weiler2016millimeter}. Considering surfaces as quasi-perfect mirrors, when a user is not reachable via a direct beam the cell search algorithm can leverage ray reflections to provide the user with a minimum level of power. In the ideal scenario where the network is perfectly aware of users' locations and obstacles' shapes and positions, the cell discovery procedure could promptly compute beamforming parameters at BS and MT side for the best reflected path. However, a detailed map of the area may be unpractical to retrieve, or it may require time-consuming in-field surveys. A more practical approach is based on geo-located context Data Bases (DBs). 

Obstacles may notably delay the cell discovery process, in particular when advanced discovery algorithms are implemented. Indeed, algorithms are designed to focus BS beams around the nominal user's position, assuming LOS conditions. However, the reflected path can substantially deviate from the LOS path, thus a long sequence of candidate BS beams could be explored before finding the one covering the user. At MT side, the impact is less relevant. Indeed, the simple sequential beam scanning starts from a random position and it less influenced by LOS path deviations.

The effect of obstacles can be summarized as an error shift between the direct BS beam covering the nominal user's position and the correct BS beam used to reach the user in the real position. This shift presents similar features as the position error. Nevertheless, it is deterministic, allowing the algorithm, once the user is discovered, to store the correct beamforming configuration in the DB in terms of \{$(x,y),w,d$\} tuples, where user's position $(x,y)$ is bonded to BS beam parameters (beam width $w$ and beam direction $d$). This approach takes advantage of past attempts, speeding up the discovery process when a new user service request comes from a ``known'' position (or close to that). 

\begin{figure}[t!]
\centering
\vspace{-1cm}
\includegraphics[width=\columnwidth]{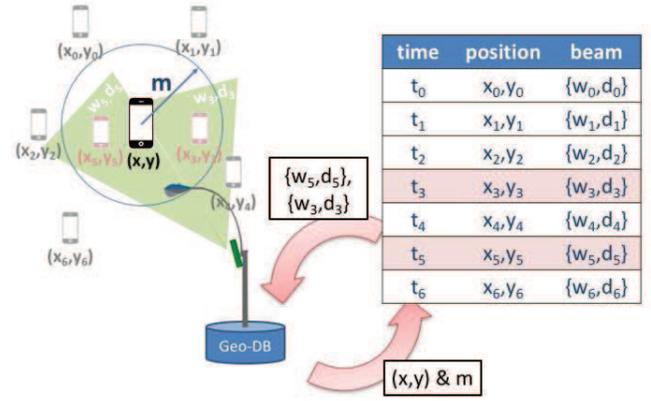}
\vspace{-1.2cm}
\caption{Example of range $m$ setup in the geo-located context database.}
\label{fig:dbrange}
\end{figure}

The use of the DB information, however, is not straightforward: Here we carefully address the definition of ``closeness'' to a position associated to a DB entry. To this aim, we introduce a range $m$. Given a new user requesting access from a nominal position $p=(x',y')$, $m$ indicates the maximum distance from $p$ within which candidate entries can be considered in the DB, as shown in Fig.~\ref{fig:dbrange}. The discovery phase tests all BS beam configurations of valid entries, which are sorted based on their distance from $p$. Note that, although multiple entries can suggest the same beam configuration, each beam configuration is activated only once, at the closest entry. If the user is not found using DB information, the search proceeds by activating one of the algorithms proposed in Section~\ref{sec:obstaclefree_procedures}.

The range $m$ is strictly related to the spatial correlation of the scenario with obstacles. If $m$ is very small, the DB information may be underutilized. Only new mobile terminals placed very close to a known position can benefit of past attempts, even if the same beam configuration could be used for farther users as well. Vice versa, very large ranges may introduce misleading information. Indeed, the beam configurations that can cover very far-away positions are in general considerably different, therefore relying on ``far'' DB entries will likely produce unsuccessful attempts. In addition, a further source of spatial incoherence is the obstacle shadowing effect, which can change the successful beam configuration even for relatively close positions. Therefore, a good trade-off for designing $m$ must be investigated. We discuss these issues and provide a numerical evaluation in the next section.

\subsection{Performance evaluation}
% !TEX root = main.tex

We carried out an exhaustive simulation campaign to evaluate the impact of the geo-located context DB in scenarios with obstacles. Numerical results are obtained with our MATLAB simulator considering the above-mentioned scenarios (please refer to Section~\ref{sec:obstaclefree_results}), unless differently specified.

A $450 m$ x $350 m$ deployment area is characterized by $20 m$ x $20 m$ square obstacles uniformly dropped. We consider obstacles as opaque bodies with reflecting edges, we assume the area boundary reflecting as well. Since we investigate a simplified 2D scenario, ground reflections are statistically included in the propagation model in terms of random fading, i.e., they are not considered as possible reflected paths to reach MTs.

While the propagation model for LOS paths is the same as in Section~\ref{sec:obstaclefree_results}, we model the path loss of reflected paths (by approximating it to the first and unique reflection) with the following equation:
\begin{equation}
\text{P}_R = 82.02 + k\cdot 10 \log\left(\frac{l_r}{l_0}\right) + R + F
\end{equation}
where the reference distance $l_0$ is $5$ meters and $k$ is $2.36$, if the distance between the transmitter and the receiver is longer than the reference distance, or $2.00$ otherwise. Reflection losses are summarized by:
\begin{eqnarray*}
&R = 20 \log \left( \frac{\sin \theta - \sqrt{B}}{sin \theta + \sqrt{B}} \right),\, B = \epsilon - \cos^2 \theta;\\
&F = \frac{-80}{\ln 10}\left(\frac{\pi \sigma \sin \theta}{\lambda}\right)^2.
\end{eqnarray*}
We consider $\epsilon$ and $\sigma$ as roughness and reflection coefficients of the material (namely, $\sigma = 0.2 mm$ and $\epsilon = 4 + 0.2j$ \cite{miwebaD51}), while $\theta$ and $l_r$ as the reflection angle and the length of the reflected path, respectively.

\begin{figure} [t!]
\centering
	\includegraphics[width=0.45\textwidth]{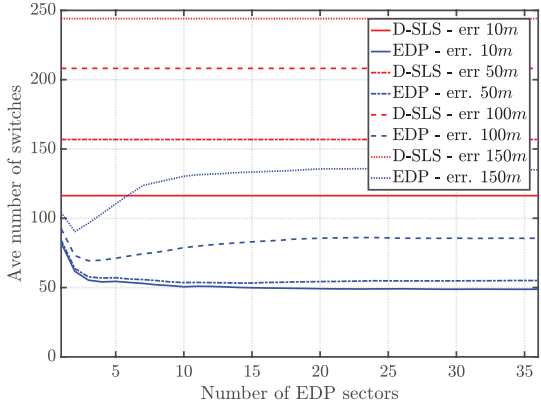}
	\caption{\footnotesize Performance comparison of D-SLS EDP algorithm at different location error values when varying the number of EDP sectors.}
	\label{fig:obst_EDPsectors}
\end{figure}

In Fig.~\ref{fig:obst_EDPsectors} we first investigate the impact of the variation of the number of sectors $n$ (or, equivalently, the sector width $2\pi/n$) on the performance of EDP algorithm in a scenario where the DB is not available. The figure indicates the average number of antenna configuration switches before a connection is established using EDP, in case of four different location error values. Results of D-SLS algorithm are reported as well for the sake of comparison. In case of a location error of $10m$, we can see a remarkable delay decrease when the number of sectors goes from $1$ to $2$, then, the improvement dramatically reduces further increasing the number of sectors. The behavior of the $50m$-error curve is similar to the $10m$-error one, except for a weak minimum around $12$-$13$ sectors. The presence of a minimum is more evident when the location error increases, in addition, the minimum moves left, towards smaller numbers of sectors. The reason is explained by the following trade-off. On the one hand, a few and wide EDP sectors increases the rendezvous time because of the delayed exploration of areas far from the BS, i.e., longer reflected paths requires narrower beams to be covered. On the other hand, several and narrow sectors increase the number of required switches as well, due to the higher sensibility to location errors caused by the increased angular-selectivity. When the number of sectors excessively increases, large location errors produce an adverse effect and the finer context exploitation of the EDP algorithm is completely canceled by the sensibility to location errors.
Interestingly, this behavior does not occur in an obstacle-free scenario as the LOS path is always available, thereby minimizing the benefit of quickly exploring far-away areas.

\begin{figure} [t!]
\centering
	\subfigure[No location error.]
	{
	\includegraphics[width=0.45\textwidth]{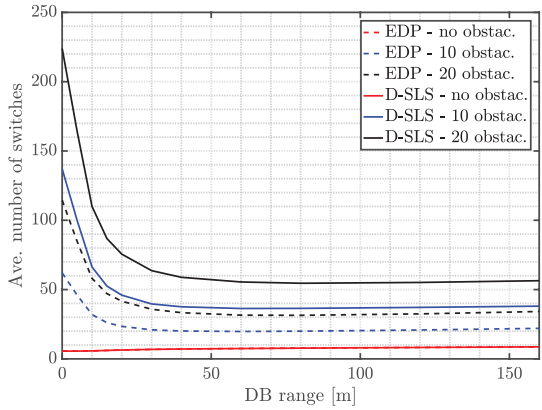}
	\label{fig:switches_vs_range_noerr}
	}
	\subfigure[$80m$ location error.]
	{
	\includegraphics[width=0.45\textwidth]{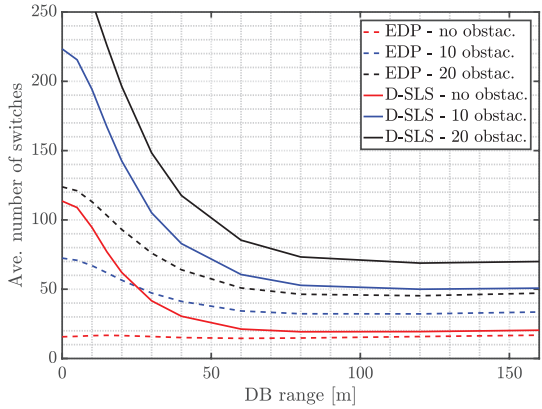}	
	\label{fig:switches_vs_range_err80}
	}

	\caption{\footnotesize Performance evaluation of DB-aided algorithms in terms of average number of beam switches varying location error and number of obstacles in the area. 
	}
	\label{fig:obstacle_algorithms_comp}
\end{figure}

Considering the DB, Fig.~\ref{fig:obstacle_algorithms_comp} shows the performance in terms of number of MT switches required to connect incoming users, for different values of user location error and different numbers of obstacles dropped into the area. The x-axis of the plots shows the variation of range $m$ used to process DB entries and the value $m=0$ indicates the DB is not used.

In Fig.~\ref{fig:switches_vs_range_noerr}, wherein a perfect location information is assumed, the presence of obstacles increases the number of required switches. Indeed, a BS beam directed to the nominal user's location cannot reach the user at his real position. This triggers an antenna configuration search. Clearly, more obstacles lead to a higher average number of antenna switches needed to establish a connection. Indeed, augmenting the number of obstacles directly increases the probability that the LOS beam cannot cover the mobile terminal. In addition, more obstacles make the difference between the direct beam and the covering beam larger, thus, the search for a new antenna configuration longer. The effect of the DB is to avoid this search after the first successful access. Accordingly, the figure shows how the improvement obtained by using the DB information is larger when more obstacles are dropped into the area. 

A different scenario is evaluated in Fig.~\ref{fig:switches_vs_range_err80}, where a large location error of $80m$ is considered. However, the advantages of using the DB information are still evident, the performance gain is even greater. A first aspect to note is that, as expected, the number of required switches is much higher than in the case of no location error, due to the combined effect of obstacles and location error. We can also note that the impact of the DB is more evident when D-SLS is used. Indeed, since the D-SLS is less efficient than EDP, finding a user with the DB information allows to save a larger number of beam switches.

\begin{figure} [t!]
\centering
	\includegraphics[width=0.45\textwidth]{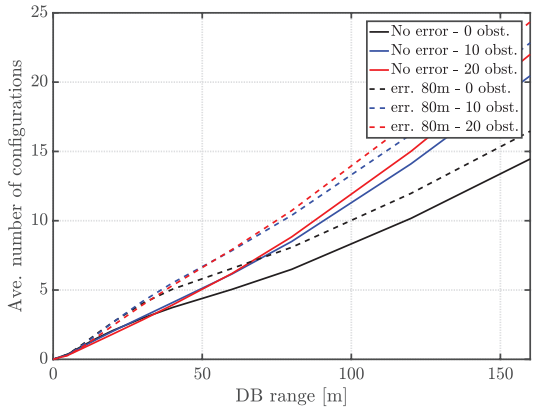}
	\caption{\footnotesize Average number of DB candidate configurations varying location error and number of obstacles in the area.
	}
	\label{fig:configuration_comp}
\end{figure}

\begin{figure} [t!]
\centering
	\includegraphics[width=0.45\textwidth]{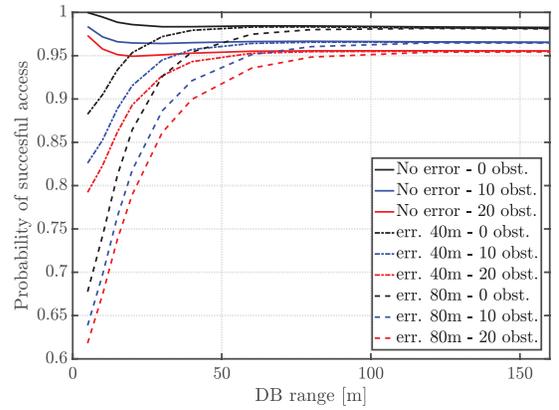}
	\caption{\footnotesize Probability of successful memory access varying location error and number of obstacles in the area.
	}
	\label{fig:success_comp}
\end{figure}

The impact of the range $m$ on the performance can be understood from Fig.~\ref{fig:configuration_comp} and Fig.~\ref{fig:success_comp}. Fig.~\ref{fig:configuration_comp} shows the average number of antenna configurations available in the DB for users that eventually connect to the BS using only DB information. It analyzes the dependence of this number on the range $m$ in scenarios with different numbers of dropped obstacles and location error intensities. As expected, it clearly shows that a larger DB range provides more configurations to be tested for incoming users. Note that D-SLS and EDP values are the same. Since both algorithms get the same successful beam, the DB will be populated with the same content as well. 

In addition, Fig.~\ref{fig:configuration_comp} shows that as the number of obstacles increases, users discovered only using DB information have seen a larger number of antenna configurations. The reason lies in the fact that obstacles harshen the environment and increase selectivity, so that only users with many candidate beams in the DB can successfully find the best one. Moreover, applying the geo-located DB in scenarios with higher location errors produces larger sets of candidate beam configurations. Indeed, due to the higher location error, a certain DB range, which selects DB entries according to their nominal positions, will correspond to a sparser set of real positions. Those real positions are reached by a larger variety of beam configurations with respect to the case there was no location error, i.e., they were close to the nominal ones.

Fig.~\ref{fig:success_comp} shows the conditioned probability of successfully establishing a connection using the DB information, provided that at least one candidate DB entry is available and considered for the incoming user. The impact of the range $m$ on the success probability is the result of a trade-off between two aspects. Increasing $m$ causes users ``relatively far'' from DB entries to have non-empty candidate sets. This generally decreases the conditioned success probability because it brings in users ``relatively badly positioned'' with respect to DB entries. However, increasing $m$ enlarges the set of beam configurations that can be tested when the DB is accessed, this increases the success probability, especially in case of location errors. The final emerging behavior depends on the actual numbers of obstacles and on the location error intensity, determining the curves' trend in the figure.

Interestingly enough, in the ideal scenario with no obstacles and no location error described in Fig.~\ref{fig:switches_vs_range_noerr} (overlapping red curves), the use of the DB is detrimental. However, this ideal scenario does not require the presence of a geo-located context DB; even a cell discovery algorithm is redundant. In fact, with no obstacles and perfect accuracy, users can be perfectly reached by directly beamforming on their nominal positions. Moreover, note that the curves in the figure have a week minimum point, after that, further increasing $m$ slightly increases the number of required steps. In fact, as stated in the previous paragraph, enlarging the range makes the information of selected beam candidates be poorly correlated with users' positions, thus resulting in a wastage of unsuccessful beamforming attempts.

\section{Implementation aspects}
\label{sec:implementability}
% !TEX root = main.tex
Current cellular deployments are characterized by the lack of multi-connectivity solutions able to orchestrate and fully exploit different Radio Access Technologies to provide context-based throughput improvements. In order to address this issue, an approach based on BS-assisted networks, also known as ``Phantom Cell'' approach~\cite{phantom_cell}, has been recently proposed as a solution offering a continuous service by means of the BS (or eNodeB, according to the 3GPP terminology) coverage, while capitalizing the ultra capacity access such that provided by mm-wave small cells, as previously presented in Section~\ref{sec:introduction}. 

The Phantom Cell concept relies on the C-plane/U-plane splitting. While the C-plane is provided by eNodeBs (eNBs) for all users within the network, the U-plane is provided by mm-wave small cells only for those users within the ultra-capacity coverage. Hence, the radio resource control (RRC) connection phase between MTs (or UEs, according to the 3GPP terminology) and the mm-wave small cells, e.g., channel establishment and release operations, are directly controlled by the eNB. Although mm-wave small cells do not provide a cell identification signals, they might be directly accessed through the eNB signaling. This benefits the control signaling in terms of handover procedures, which are significantly reduced. However, it requires low-latency connections, such as optical fibers, between the eNB and the mm-wave small cells. This architecture enables the use of mm-wave technologies in the radio access networks.

%Those architectural issues are out of the scope of this paper.

\begin{figure}[hbtp]
\centering
\vspace{-0.2cm}
\includegraphics[clip,width=0.5\textwidth]{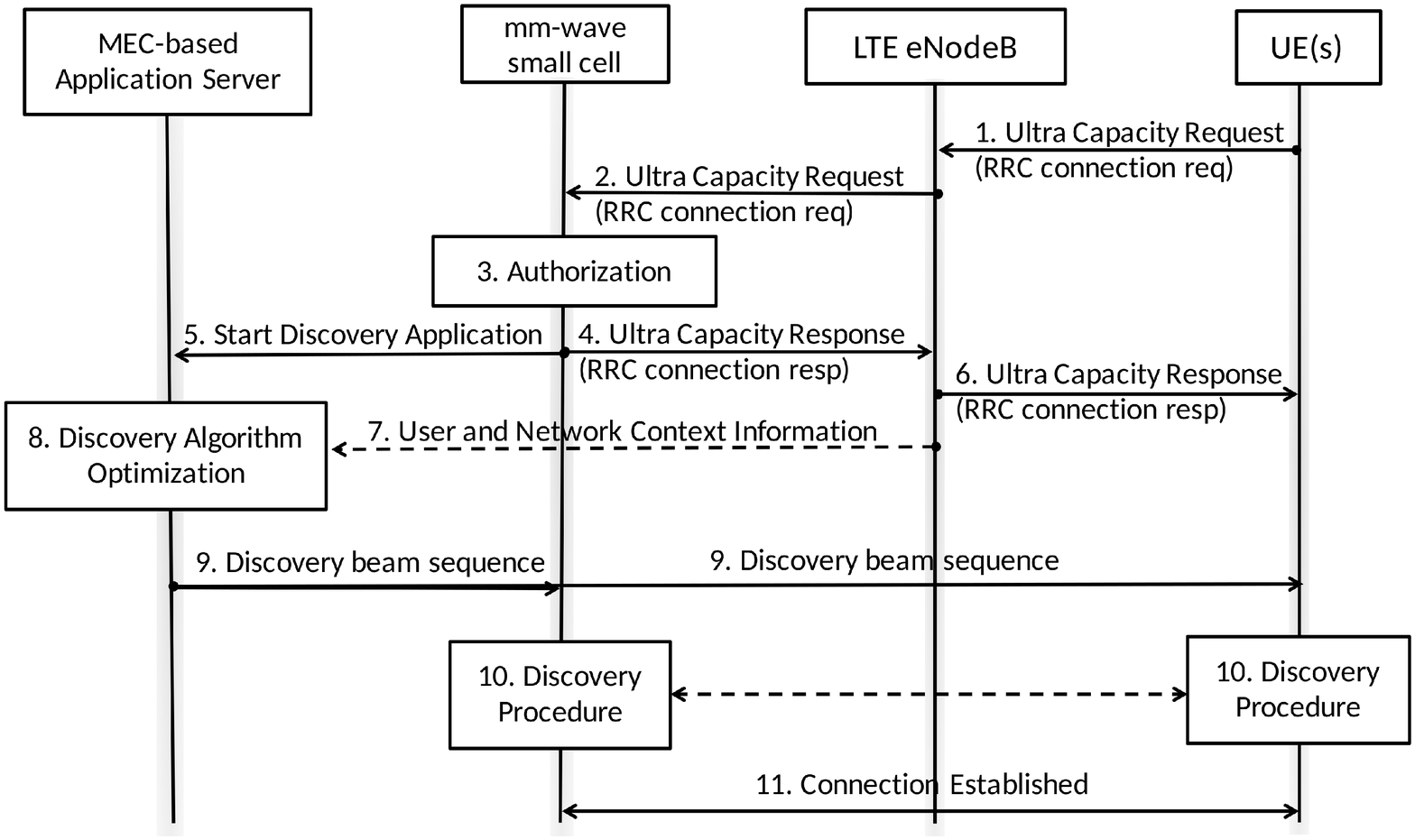}
\vspace{-5mm}
\caption{Flow diagram for Cell Discovery Procedure in 3GPP architecture.}
\label{fig:act_diagram}
\end{figure}

When a user crosses an ultra capacity area, it might be granted to access to the mm-wave small-cell network, as depicted in Fig.~\ref{fig:act_diagram}. The user sends the ultra capacity connection request (1.) through the control plane. The eNodeB activates the new connection procedure by issuing the ultra capacity requests towards best candidate mm-wave small cells that can cover the user (2.). After checking the resource availability, the mm-wave cell accepts the new user connection (3.). The discovery procedure starts afterwards (5.) and, once the user is discovered, the ultra capacity connection between the mm-wave small cell and the user can be correctly established (11.). The figure also shows how this system can be fully supported by a new ICT standardization framework aiming at providing more ``intelligence'' towards the network edge, namely the {\it Mobile Edge Computing} platform~\cite{mec_white_paper}.

%\subsection{Mobile Edge Computing}

Mobile Edge Computing (MEC) is a ISG (Industry Specification Group) defined by the ETSI standardization organization, which started in 2014 from a consortium of the major industrial partners and mobile operators. The main objective of MEC is to provide an IT service environment at the edge of the network, wherein user-specific applications may be developed. The main advantage relies on the very low latency and high bandwidth which facilitate a real-time access to radio resources. The MEC platform might be placed at base stations as well as in data-centers close to the access network, to provide processing and data storage functionalities. The standardization group has the goal of defining advanced APIs, based on both network and user context, to develop network applications for increasing the user quality of experience, such as augmented reality or gaming.

In our perspective, MEC platform enables the improvement of network management operations as well, by allowing the design of advanced functionalities, computational-intensive and/or storage-demanding, operating on a local basis. Supporting mm-wave small cells with a MEC platform allows network operators to develop application-specific discovery algorithms, e.g., SLS, D-SLS or EDP, to beam the users and provide ultra-capacity connectivity. Applications can also coordinate the activity of a pool of small cells and process large DBs of context information, gathered at different places in different times, in order to speed up the discovery.

As shown in Fig.~\ref{fig:act_diagram}, when an ultra-capacity connection request is received, the mm-wave small cell requests the activation of an instance of the Discovery Application on the MEC-based Application Server (5.), the application runs in a virtualized environment and collects user and network information via the MEC API in order to optimize the cell discovery process (7.). After the customization of the discovery phase parameters according to the specific context information (8.), the Discovery Application dispatches the best sequences of beam configuration identifiers to both small cells and users, through the C-plane connection (9.). Scanning through the provided sequences, the mm-wave small cell and the user eventually beam each other. Then, the Mobile Management Entity (MME) can starts the soft-handover: the Serving Gateway (S-GW) will instantiate a new GTP tunnel with the user through the mm-wave small cell. The ultra-capacity connection is successfully established (11.).

\section{Concluding remarks}
\label{sec:concl}
% !TEX root = main.tex

We have analyzed the directional cell discovery problem in mm-wave 5G networks characterized by the C-/U-plane split. We have shown that currently available solutions are not suitable for mm-wave highly-directional systems and novel approaches must be designed.

We have proposed new discovery algorithms enhanced by the context-information available through the separated C-plane link. Results show that the performance of the algorithms depends on user distribution, however, they greatly outperform conventional SLS procedures, even in case of low information accuracy. In addition, we have developed an analytic model that predicts the cell discovery time with very high accuracy at both BS and MT side.

We have also investigated the beneficial impact of a geo-located context database, which can substantially improve the performance of the proposed algorithm in case of obstacles, basically enabling mm-wave cell discovery in future network scenarios.

\ifCLASSOPTIONcompsoc
  \section*{Acknowledgments}
\else
  \section*{Acknowledgment}
\fi

The research leading to these results has been partially supported by the EU 7th Framework Program (FP7-ICT-2013-EU-Japan) under grant agreement number 608637 (MiWEBA).

\ifCLASSOPTIONcaptionsoff
  \newpage
\fi

\bibliographystyle{IEEEtran}
\bibliography{bibliography}

\begin{thebibliography}{10}
\providecommand{\url}[1]{#1}
\csname url@samestyle\endcsname
\providecommand{\newblock}{\relax}
\providecommand{\bibinfo}[2]{#2}
\providecommand{\BIBentrySTDinterwordspacing}{\spaceskip=0pt\relax}
\providecommand{\BIBentryALTinterwordstretchfactor}{4}
\providecommand{\BIBentryALTinterwordspacing}{\spaceskip=\fontdimen2\font plus
\BIBentryALTinterwordstretchfactor\fontdimen3\font minus
  \fontdimen4\font\relax}
\providecommand{\BIBforeignlanguage}[2]{{%
\expandafter\ifx\csname l@#1\endcsname\relax
\typeout{** WARNING: IEEEtran.bst: No hyphenation pattern has been}%
\typeout{** loaded for the language `#1'. Using the pattern for}%
\typeout{** the default language instead.}%
\else
\language=\csname l@#1\endcsname
\fi
#2}}
\providecommand{\BIBdecl}{\relax}
\BIBdecl

\bibitem{survey_rappaport}
S.~Rangan, T.~Rappaport, and E.~Erkip, ``Millimeter-wave cellular wireless
  networks: Potentials and challenges,'' \emph{Proceedings of the IEEE}, vol.
  102, no.~3, pp. 366--385, March 2014.

\bibitem{HHI}
A.~Maltsev, A.~Pudeyev, I.~Karls, I.~Bolotin, G.~Morozov, R.~Weiler, M.~Peter,
  and W.~Keusgen, ``Quasi-deterministic approach to mmwave channel modeling in
  a non-stationary environment,'' in \emph{IEEE Globecom 2014, Workshop on
  Emerging Technologies for 5G Wireless Cellular Networks}, 2014.

\bibitem{mmwave_eLAN}
A.~Ghosh, T.~Thomas, M.~Cudak, R.~Ratasuk, P.~Moorut, F.~Vook, T.~Rappaport,
  G.~Maccartney, S.~Sun, and S.~Nie, ``Millimeter-wave enhanced local area
  systems: A high-data-rate approach for future wireless networks,''
  \emph{Selected Areas in Communications, IEEE Journal on}, vol.~32, no.~6, pp.
  1152--1163, June 2014.

\bibitem{coverage_analysis}
T.~Bai and R.~Heath, ``Coverage and rate analysis for millimeter-wave cellular
  networks,'' \emph{Wireless Communications, IEEE Transactions on}, vol.~14,
  no.~2, pp. 1100--1114, Feb 2015.

\bibitem{bcgwons}
A.~Capone, A.~F. dos Santos, I.~Filippini, and B.~Gloss, ``Looking beyond green
  cellular networks,'' in \emph{Wireless On-demand Network Systems and Services
  (WONS), 2012 9th Annual Conference on}.\hskip 1em plus 0.5em minus
  0.4em\relax IEEE, 2012, pp. 127--130.

\bibitem{SDN-mmwaves}
Y.~Niu, Y.~Li, M.~Chen, D.~Jin, and S.~Chen, ``A cross-layer design for a
  software-defined millimeter-wave mobile broadband system,'' \emph{IEEE
  Communications Magazine}, vol.~54, no.~2, pp. 124--130, February 2016.

\bibitem{split}
R.~Weiler, W.~Keusgen, I.~Filippini, and A.~Capone, ``Split control plane
  functionality in millimeter-wave overlay access,'' in \emph{1st International
  Conference on 5G for Ubiquitous Connectivity}, 2014.

\bibitem{NFV-specs}
``{ETSI ISG for NFV, ETSI GS NFV 002: Network Functions Virtualisation (NFV);
  Architectural Framework},'' January 2015.

\bibitem{li2013anchor}
Q.~C. Li, H.~Niu, G.~Wu, and R.~Q. Hu, ``Anchor-booster based heterogeneous
  networks with mmwave capable booster cells,'' in \emph{Globecom Workshops (GC
  Wkshps)}.\hskip 1em plus 0.5em minus 0.4em\relax IEEE, 2013, pp. 93--98.

\bibitem{raghavan2016beamforming}
V.~Raghavan, J.~Cezanne, S.~Subramanian, A.~Sampath, and O.~Koymen,
  ``Beamforming tradeoffs for initial {UE} discovery in millimeter-wave {MIMO}
  systems,'' \emph{IEEE Journal of Selected Topics in Signal Processing},
  vol.~10, no.~3, pp. 543--559, 2016.

\bibitem{rappaport2012angle}
M.~Akdeniz, Y.~Liu, M.~Samimi, S.~Sun, S.~Rangan, T.~Rappaport, and E.~Erkip,
  ``Millimeter wave channel modeling and cellular capacity evaluation,''
  \emph{Selected Areas in Communications, IEEE Journal on}, vol.~32, no.~6, pp.
  1164--1179, June 2014.

\bibitem{weiler2016millimeter}
R.~Weiler, W.~Keusgen, A.~Maltsev, T.~Kuhne, A.~Pudeyev, L.~Xian, J.~Kim, and
  M.~Peter, ``Millimeter-wave outdoor access shadowing mitigation using
  beamforming arrays,'' in \emph{Antennas and Propagation (EuCAP), 2016
  European Conference on}, April 2016, pp. 4568--4573.

\bibitem{jakllari2007integrated}
G.~Jakllari, W.~Luo, and S.~V. Krishnamurthy, ``An integrated neighbor
  discovery and {MAC} protocol for ad hoc networks using directional
  antennas,'' \emph{Wireless Communications, IEEE Transactions on}, vol.~6,
  no.~3, pp. 1114--1024, 2007.

\bibitem{choudhury2006on}
R.~Choudhury, X.~Yang, R.~Ramanathan, and N.~Vaidya, ``On designing mac
  protocols for wireless networks using directional antennas,'' \emph{Mobile
  Computing, IEEE Transactions on}, vol.~5, no.~5, pp. 477--491, May 2006.

\bibitem{Ramanathan2005ad}
R.~Ramanathan, J.~Redi, C.~Santivanez, D.~Wiggins, and S.~Polit, ``Ad hoc
  networking with directional antennas: a complete system solution,''
  \emph{Selected Areas in Communications, IEEE Journal on}, vol.~23, no.~3, pp.
  496--506, March 2005.

\bibitem{Korakis2008cdr-mac}
T.~Korakis, G.~Jakllari, and L.~Tassiulas, ``{CDR-MAC}: A protocol for full
  exploitation of directional antennas in ad hoc wireless networks,''
  \emph{Mobile Computing, IEEE Transactions on}, vol.~7, no.~2, pp. 145--155,
  Feb 2008.

\bibitem{Choudhury2004deafness}
R.~Choudhury and N.~Vaidya, ``Deafness: a {MAC} problem in ad hoc networks when
  using directional antennas,'' in \emph{Network Protocols, 2004. ICNP 2004.
  Proceedings of the 12th IEEE International Conference on}, Oct 2004, pp.
  283--292.

\bibitem{wang2009beam}
J.~Wang, Z.~Lan, C.-W. Pyo, T.~Baykas, C.-S. Sum, M.~A. Rahman, J.~Gao,
  R.~Funada, F.~Kojima, H.~Harada \emph{et~al.}, ``Beam codebook based
  beamforming protocol for multi-{Gbps} millimeter-wave {WPAN} systems,''
  \emph{Selected Areas in Communications, IEEE Journal on}, vol.~27, no.~8, pp.
  1390--1399, 2009.

\bibitem{singh2009blockage}
S.~Singh, F.~Ziliotto, U.~Madhow, E.~Belding, and M.~Rodwell, ``Blockage and
  directivity in 60 {GHz} wireless personal area networks: from cross-layer
  model to multihop {MAC} design,'' \emph{Selected Areas in Communications,
  IEEE Journal on}, vol.~27, no.~8, pp. 1400--1413, 2009.

\bibitem{chandra2014adaptive}
K.~Chandra, R.~Prasad, I.~Niemegeers, and A.~Biswas, ``Adaptive beamwidth
  selection for contention based access periods in millimeter wave {WLANs},''
  in \emph{Consumer Communications and Networking Conference (CCNC), 2014 IEEE
  11th}, Jan 2014, pp. 458--464.

\bibitem{chen2013directional}
Q.~Chen, J.~Tang, D.~Wong, X.~Peng, and Y.~Zhang, ``Directional cooperative
  {MAC} protocol design and performance analysis for {IEEE} 802.11ad {WLANs},''
  \emph{Vehicular Technology, IEEE Transactions on}, vol.~62, no.~6, pp.
  2667--2677, July 2013.

\bibitem{beam_searching}
H.~Shokri-Ghadikolaei, L.~Gkatzikis, and C.~Fischione, ``Beam-searching and
  transmission scheduling in millimeter wave communications,'' in
  \emph{Communications (ICC), 2015 IEEE International Conference on}, June
  2015, pp. 1292--1297.

\bibitem{beam_steering_alg}
A.~Patra, L.~Simi\'{c}, and P.~M\"{a}h\"{o}nen, ``Smart mm-wave beam steering
  algorithm for fast link re-establishment under node mobility in 60 {GHz}
  indoor {WLANs},'' in \emph{Proceedings of the 13th ACM International
  Symposium on Mobility Management and Wireless Access (MobiWac)}, 2015, pp.
  53--62.

\bibitem{capone2015context}
A.~Capone, I.~Filippini, and V.~Sciancalepore, ``Context information for fast
  cell discovery in mm-wave {5G} networks,'' in \emph{European Wireless 2015,
  21th European Wireless Conference}, May 2015.

\bibitem{capone2015_2}
A.~Capone, I.~Filippini, V.~Sciancalepore, and D.~Tremolada, ``Obstacle
  avoidance cell discovery using mm-wave directive antennas in {5G} networks,''
  in \emph{IEEE PIMRC 2015, Workshop on Cloud Cooperated Heterogeneous Cellular
  Networks for 5G}, Aug 2015.

\bibitem{dir_cell_disc_rappaport}
C.~Barati, S.~Hosseini, S.~Rangan, P.~Liu, T.~Korakis, S.~Panwar, and
  T.~Rappaport, ``Directional cell discovery in millimeter wave cellular
  networks,'' \emph{Wireless Communications, IEEE Transactions on}, vol.~14,
  no.~12, pp. 6664--6678, Dec 2015.

\bibitem{mac_layer_perspective}
H.~Shokri-Ghadikolaei, C.~Fischione, G.~Fodor, P.~Popovski, and M.~Zorzi,
  ``Millimeter wave cellular networks: A {MAC} layer perspective,''
  \emph{Communications, IEEE Transactions on}, vol.~63, no.~10, pp. 3437--3458,
  Oct 2015.

\bibitem{directional_access}
C.~Barati, S.~Hosseini, S.~Rangan, P.~Liu, T.~Korakis, and S.~Panwar,
  ``Directional cell search for millimeter wave cellular systems,'' in
  \emph{Signal Processing Advances in Wireless Communications (SPAWC), 2014
  IEEE 15th International Workshop on}, June 2014, pp. 120--124.

\bibitem{zorzi1}
M.~Giordani, M.~Mezzavilla, C.~Barati, S.~Rangan, and M.~Zorzi, ``Comparative
  analysis of initial access techniques in {5G} mm-wave cellular networks,'' in
  \emph{50th Annual Conference on Information Sciences and Systems (CISS), Mar.
  2016}.

\bibitem{zorzi2}
W.~Abbas and M.~Zorzi, ``Context information based initial cell search for
  millimiter wave {5G} cellular networks,'' available on arXiv:1605.019930v, 6
  May 2016, Tech. Rep., 2016.

\bibitem{redondi2013context}
A.~Redondi, I.~Filippini, and A.~Capone, ``Context management in
  energy-efficient radio access networks,'' in \emph{Digital
  Communications-Green ICT (TIWDC), 2013 24th Tyrrhenian International Workshop
  on}.\hskip 1em plus 0.5em minus 0.4em\relax IEEE, 2013, pp. 1--5.

\bibitem{miwebaD51}
\BIBentryALTinterwordspacing
``{FP7-ICT-608637 MiWEBA Project Deliverable D5.1 - Channel Modeling and
  Characterization},'' June 2014. [Online]. Available:
  \url{http://www.miweba.eu/wp-content/uploads/2014/07/MiWEBA_D5.1_v1.01.pdf}
\BIBentrySTDinterwordspacing

\bibitem{GPS_book}
A.~P. Andrews, L.~R. Weill, and M.~S. Grewal, \emph{Global Positional Systems,
  Inertial Navigation, and Integration}.\hskip 1em plus 0.5em minus 0.4em\relax
  John Wiley \& Sons, Inc. 2001.

\bibitem{phantom_cell}
H.~ISHII, Y.~Kishiyama, and H.~Takahashi, ``A novel architecture for {LTE-B}:
  {C}-plane/{U}-plane split and phantom cell concept,'' in \emph{Globecom
  Workshops (GC Wkshps)}, Dec 2012, pp. 624--630.

\bibitem{mec_white_paper}
\BIBentryALTinterwordspacing
``{Mobile-Edge Computing},'' September 2014. [Online]. Available:
  \url{https://portal.etsi.org/Portals/0/TBpages/MEC/Docs/Mobile-edge_Computing_-_Introductory_Technical_White_Paper_V1%2018-09-14.pdf}
\BIBentrySTDinterwordspacing

\end{thebibliography}


%\end{IEEEbiography}


%\newpage



%\vfill


\end{document}